%% file: 00_Main.tex
\documentclass[journal]{IEEEtran}
\ifCLASSINFOpdf
\else
\fi
\ifCLASSOPTIONcompsoc
 \usepackage[caption=false,font=normalsize,labelfont=sf,textfont=sf]{subfig}
\else
 \usepackage[caption=false,font=footnotesize]{subfig}
\fi
\usepackage{siunitx}
\usepackage[utf8]{inputenc}	
\usepackage{tabularx}
\usepackage{multirow,multicol,colortbl,booktabs,lipsum}
\usepackage{threeparttable}
\usepackage{booktabs}
\usepackage{amsmath}
\usepackage{bm}
\usepackage{cases}
\usepackage{empheq, nccmath}
\usepackage{amssymb}
\usepackage{array,graphicx}
\usepackage{float}
\usepackage[table,xcdraw,dvipsnames]{xcolor}
\usepackage{amssymb}
\usepackage{pifont}
\usepackage[nocomma, short]{optidef} 
\usepackage{units} 
\usepackage{subfig}
\usepackage{enumitem}  
\usepackage{dsfont}
\usepackage{amsfonts, amsthm }

\begin{document}

        \input{01_Title.tex}

        \input{02_Abstract.tex}
        \input{03_Nomenclature.tex}

        \input{04_Introduction.tex}
        \input{05_MarketModels.tex}

        \input{06_MPEC.tex}

        \input{07_Solution.tex}

        \input{08_CaseStudy}

        \input{09_Conclusion.tex}

        \bibliographystyle{IEEEtran}
        \bibliography{98_References.bib}
        
        \input{97_Appendix.tex}

\end{document}

%% file: 01_Title.tex
%
\title{Dynamic Reserve and Transmission Capacity Allocation in Wind-Dominated Power Systems}

\author{Nicola Viafora, \IEEEmembership{Student Member, IEEE,} Stefanos Delikaraoglou, \IEEEmembership{Member, IEEE,} Pierre Pinson, \IEEEmembership{Fellow, IEEE,} Gabriela Hug, \IEEEmembership{Senior Member, IEEE,} Joachim Holbøll, \IEEEmembership{Senior Member, IEEE}}

\markboth{Submitted to IEEE Transactions on Power Systems, March~2020}%
{Shell \MakeLowercase{\textit{et al.}}: Bare Demo of IEEEtran.cls for IEEE Journals}
%



\maketitle

%% file: 02_Abstract.tex
\begin{abstract}
The large shares of wind power generation in electricity markets motivate higher levels of operating reserves. However, current reserve sizing practices fail to account for important topological aspects that might hinder their deployment, thus resulting in high operating costs. Zonal reserve procurement mitigates such inefficiencies, however, the way the zones are defined is still open to interpretation. This paper challenges the efficiency of predetermined zonal setups that neglect the location of stochastic power production in the system, as well as the availability, cost and accessibility of flexible generating units. To this end, we propose a novel reserve procurement approach, formulated as a two-stage stochastic bilevel model, in which the upper level  identifies a number of contiguous reserve zones using dynamic grid partitioning and sets zonal requirements based on the total expected operating costs. Using two standard IEEE reliability test cases, we show how the efficient partitioning of reserve zones can reduce expected system cost and promote the integration of stochastic renewables.  
\end{abstract}


%
\IEEEpeerreviewmaketitle

%% file: 03_Nomenclature.tex
 \vspace{-14pt}

\section*{Nomenclature}


 \vspace{-8pt}

\subsection{Sets and Indices}
\begin{IEEEdescription}[\IEEEusemathlabelsep\IEEEsetlabelwidth{$s^+_{\ell,z}, s^-_{\ell,z} $}]
\item[$n \in \mathcal{N}$] Set of nodes.
\item[$\ell \in \mathcal{L}$] Set of transmission lines.
\item[$g \in \mathcal{G}$] Set of conventional generators.
\item[$j \in \mathcal{J}$] Set of wind power generators.
\item[$z \in \mathcal{Z}$] Set of partitions. 
\item[$s \in \mathcal{S}$] Set of scenarios.
\end{IEEEdescription}
\vspace{-0.4cm}
\subsection{Parameters}
\begin{IEEEdescription}[\IEEEusemathlabelsep\IEEEsetlabelwidth{$s^+_{\ell,z}, s^-_{\ell,z} $}]
\item[$\overline{P}_g, \underline{P}_g$] Max/min generator's output.
\item[$F_\ell$] Line rating.
\item[$R^+_g, R^-_g$] Up/down reserve capacity offer.
\item[$\Lambda^+, \Lambda^- $] Up/down deterministic reserve requirement.
\item[$C_g, C^{+/-}_g $] Generation and up/down reserve cost.
\item[$C^\text{sh}, C^\text{ct}$] Load shedding and wind curtailment cost.
\item[$\boldsymbol{H}$] Incidence matrices.
\item[$\boldsymbol{M}$] Power transfer distribution factor matrix.
\item[$D_n, \boldsymbol{D}$] Nodal load demand.
\item[$\pi_s$] Probability of scenario. 
\item[$\widehat{W}_j, W_{j,s}$] Wind power point forecast and realization.
\end{IEEEdescription}
\vspace{-0.4cm}
\subsection{Decision variables}
\begin{IEEEdescription}[\IEEEusemathlabelsep\IEEEsetlabelwidth{$s^+_{\ell,z}, s^-_{\ell,z} $}]
\item[$x_{n,z}$] Binary variable for grid partitioning.
\item[$y_{z}$] Number of nodes per zone.
\item[$\varphi_{\ell,z}$] Flowing units on line $\ell$ in zone $z$ for expressing zone connectivity.
\item[$c_{n,z}$] Root node selection.
\item[$r^+_{g,z}, r^-_{g,z} $] Up/down procured reserve.
\item[$\lambda^+_z, \lambda^-_z $] Up/down zonal reserve requirement.
\item[$\widehat{f}_\ell, f_{\ell,s}$] Expected power flow and realization in scen. \textit{s}.
\item[$p_g, \boldsymbol{p}$] Day-ahead dispatch of conventional generators.
\item[$p^+_{g,s}, p^-_{g,s}$] Up/down reserve deployment per scenario.
\item[$w_j, w^\text{ct}_{j,s}$] Scheduled and curtailed wind power.
\item[$d^\text{sh}_{n,s}$] Nodal load shedding per scenario.
\item[$w^\text{ct}_{j,s}$] Wind power curtailment per scenario.
\item[$\Gamma_\ell$] Capacity allocation margin at day-ahead market.

\end{IEEEdescription}



%% file: 04_Introduction.tex
\section{Introduction}

Several studies indicate that high shares of wind power generation require significantly more operating reserves to accommodate the uncertainty and the variability arising from forecast errors and inherent fluctuations in the wind regime \cite{Morales2009}. However, simply increasing the reserve capacity requirements does not guarantee that the system will have access to sufficient flexible resources during real-time operation, since the existing reserve capacity market is myopic about the grid topology limitations. As a result, in cases when operating reserves cannot be delivered due to network congestions, system operators have to resort to more expensive corrective actions, such as wind curtailment and load shedding.
    
An implicit way to account for network limitations during the reserve procurement process is to consider a zonal representation of the system. This approximation allows system operators to differentiate zonal reserve requirements based on expected congestion patterns and the location of stochastic power production. Nevertheless, the effectiveness of this approach is limited by the ability to define and update zone boundaries based on the operating conditions. Despite being an approximation of the true network topology, this zonal splitting approach is readily compatible with the current market structure and allows to convey to the reserve market more complete information about the balancing needs of the system at specific locations. This is a fundamental property of the more advanced energy and reserves co-optimization models based on two-stage stochastic programming \cite{Papavasiliou2013}, which however comes at the expense of violating the cost recovery and revenue adequacy properties for some uncertainty realizations \cite{Morales2014}.

This latter consideration has motivated several studies to use a stochastic bilevel programming approach that preserves the existing market structure and its desirable economic properties not only in expectation, but for every uncertainty outcome. Authors in \cite{Morales2014} adopt this framework for optimally dispatching wind power in an energy-only market, whereas authors in \cite{Dvorkin2019} employ an analogous approach to define the optimal reserve requirements in view of wind power uncertainty. In a similar vein, \cite{Delikaraoglou2019} extends this model to account for the allocation of cross-border transmission capacity between energy and reserves. Although these models have shown to improve the total expected cost in a sequential market-clearing architecture, they still lack the ability to optimally position reserves in the system, as the ideal stochastic model does. This stems from the merit-order principle enforced by the existing market design, which restricts the procurement of reserves from the cheapest generators, regardless of their location in the system. 
 
The aforementioned studies considered either a single zone or a predefined zonal setup for the reserve procurement. This paper proposes a novel Zonal Preemptive methodology, where not only zonal reserve requirements are defined, but the zone boundaries themselves are considered as decision variables. The goal is to improve the positioning of reserves in the system, while remaining compatible with the current market structure. In this work, we build upon \cite{Dvorkin2019} and \cite{Delikaraoglou2019} and we embed grid partitioning algorithms in the stochastic bilevel problem in order to identify a number of zonal reserve markets to be cleared independently. Grid partitioning algorithms have been used already in power system research for intentional islanding studies in \cite{Fan2012}-\cite{Golari2014}. However, to the best of our knowledge, this is the first attempt to rely on them for setting zonal reserve requirements. The proposed approach can be used as a decision-support tool for the grid operators for the redefinition of  reserve zones based on the location of stochastic power production, cost and expected accessibility of flexible generators' reserve capacity. 
    
While zonal reserve allocation is not a novel concept, the way the zones are defined is still open to interpretation. Existing studies base the partitioning of the system on heuristic methods that consider: active and reactive power flow sensitivities \cite{Kumar2004}; data-driven clustering techniques \cite{Wang2012}; weighted power transfer distribution factors (PTDFs) \cite{Wang2015}; reserve market clearing prices \cite{Chen2014} or simply use pre-defined partitions \cite{Khatir2013} that can be based on geographical boundaries or ownership. The proposed approach relies instead on a partitioning scheme that is solely driven by the total expected costs, thus without the need of any additional metric. 

We extend our formulation to include the ability to exchange reserve between neighbouring zones as in \cite{Delikaraoglou2019}. Setting aside part of the transmission capacity for reserve accessibility has shown to lower the total operating costs \cite{Gebrekiros2013}. However, as the zones are defined dynamically, so are the cross-zonal lines eligible for reserve sharing. The proposed methodology addresses this issue by adapting the grid partitioning constraints accordingly. Simulation results are showcased based on both IEEE RTS-24 and IEEE RTS-96 systems, where we benchmark our methodology against a sequential approach, the stochastic energy and reserve co-optimization and the stochastic bilevel with a single or predefined zones.

The remainder of the paper is organized as follows. Various reserve procurement and dispatch models are reviewed in Section \ref{sec:markets}, the proposed model is explained in details in Section \ref{sec:RCAM} whereas a solution methodology is presented in Section \ref{sec:solution}. Lastly, Section \ref{sec:results} elaborates on selected simulation results and Section \ref{sec:conclusion} draws final conclusions.

%% file: 05_MarketModels.tex
\section{Reserve Procurement and Dispatch Models} \label{sec:markets}

We first provide the mathematical formulation of the existing European market design, based on the sequential clearing of the reserve capacity, day-ahead energy and balancing markets. We then provide a compact formulation of the stochastic energy and reserve co-optimization model, emphasizing its main differences compared to the sequential approach.

\subsection{Sequential Approach}
Let $\Lambda^+$ and $\Lambda^-$ indicate the upward and downward reserve requirements. These are provided as exogenous parameters to the reserve market clearing algorithm that is formulated as 
\begin{mini!}|s|[2]                   
    {_{\Xi_\text{R}}} 
    { \mathcal{C}_\text{R} = \sum_{g \in \mathcal{G}} \left( C_g^+ r_g^+ + C_g^- r_g^- \right) }{\label{prob:RM}} 
    {} 
    \addConstraint{\sum_{g \in \mathcal{G}} r_g^+ \geq \Lambda^+, \quad }{ \sum_{g \in \mathcal{G}} r_g^- \geq \Lambda^-, }{\label{RM_rup_req}}   
    \addConstraint{0 \leq r_g^+ \leq R_g^+, \quad  }{0 \leq r_g^- \leq R_g^-,}{ \hspace{1cm}  \forall g \in \mathcal{G}, \label{RM_rup_lim}}
\end{mini!}
where $\Xi_\text{R} = \{ r_g^+, r_g^-, \forall g \}$ is the set of free variables, i.e., up- and downward reserve capacity procured from each generator. Constraints \eqref{RM_rup_req} guarantee that the pre-determined reserve requirements $\Lambda$ are met, whereas \eqref{RM_rup_lim} limit the amount of reserve that can be procured to generators' capacity offers. 

Having reserve capacity procurement $r_g^{+,*}$ and $r_g^{-,*}$ from model \eqref{prob:RM} as fixed parameters, the optimal day-ahead energy schedule for conventional $p_g$ and stochastic $w_j$ generators is obtained solving the following problem
\begin{mini!}|s|[2]                   
    {_{\Xi_\text{D}}} 
    { \mathcal{C}_\text{D} = \sum_{g \in \mathcal{G}} C_g p_g }{\label{prob:DAM}} 
    {} 
    \addConstraint{\sum_{g \in \mathcal{G}} p_g + \sum_{j \in \mathcal{J}} w_j }{= \sum_{n \in \mathcal{N}} D_n, }{\label{DAM_bal}}    
    \addConstraint{\underline{P}_g + r_g^{-,*} \leq p_g}{\leq \overline{P}_g - r_g^{+,*},}{ \hspace{0.5cm} \forall g \in \mathcal{G}, \label{DAM_p_lim}}
    \addConstraint{0 \leq w_j}{\leq \widehat{W}_j, }{ \hspace{0.5cm} \forall j \in \mathcal{J}, \label{DAM_w_lim}}
    \addConstraint{\widehat{f}_\ell}{= \boldsymbol{M}_{(\ell,\cdot)} \left( \boldsymbol{H}_\text{G}^\top \boldsymbol{p} + \boldsymbol{H}_\text{J}^\top \boldsymbol{w} - \boldsymbol{D} \right),}{\hspace{0.5cm} \forall \ell \in \mathcal{L}, \label{DAM_f}}
    \addConstraint{- F_\ell \leq \widehat{f}_\ell  }{\leq F_\ell,}{ \hspace{0.5cm} \forall \ell \in \mathcal{L} \label{DAM_f_lim} }
\end{mini!}
where $\Xi_\text{D} = \{ p_g, \forall g; \, w_j, \forall j \}$ collects the decision variables. The day-ahead power balance is enforced by constraint \eqref{DAM_bal}, whereas the production of conventional units is bounded by the minimum and maximum generation limits and procured reserves in constraint \eqref{DAM_p_lim}. Stochastic producers are assumed to be wind power generators only, whose dispatch is limited to the available point forecast $\widehat{W}_j$ in constraint \eqref{DAM_w_lim}. Employing a DC network approximation, power flows are modelled by \eqref{DAM_f} using the PTDF matrix $\boldsymbol{M}$ and are in turn restricted by the corresponding transmission capacity limits in \eqref{DAM_f_lim}. Appropriate incidence matrices $\boldsymbol{H}_\text{G}$ and $\boldsymbol{H}_\text{J}$ map conventional and stochastic generators to the respective buses in the system.

Approaching the hour of the delivery when wind power realization $W_{j,s'}$ is known, the balancing market is cleared using the following model to ensure that any deviation from the day-ahead schedule $p_g^*, w_j^*$ is balanced by appropriate re-dispatch actions for the uncertainty realization $ s = s'$. 

\begin{mini!}|s|[3]                   
    {_{\Xi_\text{B,s'}}} 
    { \mathcal{C}_{\text{B},s'} = \sum_{g \in \mathcal{G}} C_g \left( p_{g,s'}^+ - p_{g,s'}^- \right) \ldots  \\ + \sum_{j \in \mathcal{J}} C^\text{ct} w_{j,s'}^\text{ct} + \sum_{n \in \mathcal{N}} C^\text{sh} d_{n,s'}^\text{sh} \nonumber }{\label{prob:BM}} 
    {\label{prob:BM_Obj}} 
     {}{}    
    \addConstraint{ \hspace{-5.8cm} \sum_{g \in \mathcal{G}} \left( p_{g,s'}^+ - p_{g,s'}^- \right) + \sum_{j \in \mathcal{J}} \left( \Delta W_{j,s'} - w_{j,s'}^\text{ct} \right) + \sum_{n \in \mathcal{N}} d_{n,s'}^\text{sh} = 0 \label{BM_bal}}
    \addConstraint{ \hspace{-5.8cm} 0 \leq p_{g,s'}^+ }{\leq r_g^{+,*}, }{ \hspace{-2cm} \forall g \in \mathcal{G}, \label{BM_p+_lim} }
    \addConstraint{ \hspace{-5.8cm} 0 \leq p_{g,s'}^- }{\leq r_g^{-,*}, }{ \hspace{-2cm} \forall g \in \mathcal{G}, \label{BM_p-_lim} }
    \addConstraint{ \hspace{-5.8cm} 0 \leq d_{n,s'}^\text{sh} }{ \leq D_n,}{ \hspace{-2cm} \forall n \in \mathcal{N}, \label{BM_dsh} }
    \addConstraint{ \hspace{-5.8cm} 0 \leq w_{j,s'}^\text{ct} }{ \leq W_{j,s'},}{ \hspace{-2cm} \forall j \in \mathcal{J}, \label{BM_wct} }
    \addConstraint{ \hspace{-5.8cm} - F_\ell \leq f_{\ell,s'} }{\leq F_\ell,}{ \hspace{-2cm} \forall \ell \in \mathcal{L}, \label{BM_f_lim} }
    \addConstraint{ \hspace{-5.8cm} f_{\ell,s'} = \boldsymbol{M}_{(\ell,\cdot)} \big[ \boldsymbol{H}_\text{G}^\top \left( \boldsymbol{p^*} + \boldsymbol{p}_{s'}^+ - \boldsymbol{p}_{s'}^- \right) \ldots \nonumber }  
    \addConstraint{ \hspace{-5.8cm} \ldots + \boldsymbol{H}_\text{J}^\top \left( \boldsymbol{W}_{s'} - \boldsymbol{w}^\text{ct} \right) - \left( \boldsymbol{D} - \boldsymbol{d}_{s'}^\text{sh} \right) \big], }{}{ \hspace{-2cm} \forall \ell \in \mathcal{L}  \label{BM_f}}
\end{mini!}
where $\Xi_{\text{B},s'} = \{ p_{g,s'}^+, p_{g,s'}^-, \forall g; \, w_{j,s'}^\text{ct}, \forall j; \, d_{n,s'}^\text{sh}, \forall n \}$ is the set of decision variables and $\Delta W_{j,s'} = W_{j,s'} - w_j^*$ represents the system imbalance. The objective function \eqref{prob:BM_Obj} includes a cost $C_g$ for the activation of reserves from those generators that were cleared to provide reserves and have already received a capacity payment. Additionally, we assume that the grid operator has to face a cost $C^\text{ct}$ and $C^\text{sh}$ for wind power curtailment and load not supplied, respectively.
Constraint \eqref{BM_bal} is the real-time power balance, whereas constraints \eqref{BM_p+_lim}-\eqref{BM_p-_lim} limit the activation of reserve to the procured values in \eqref{prob:RM}. The use of corrective actions is limited by constraints \eqref{BM_dsh}-\eqref{BM_wct}, which model load shedding and wind curtailment, respectively. Finally, \eqref{BM_f_lim} enforce power flow limits, where real-time power flows in each scenario are modelled in \eqref{BM_f}.

\subsection{ Stochastic Energy and Reserve Co-Optimization }
An improved method based on two-stage stochastic programming allows the grid operator to jointly co-optimize reserve and energy. In this framework, the first stage models reserve as well as day-ahead energy scheduling, whereas the second stage corresponds to the balancing market under each considered realization of the uncertain variables. The two-stage stochastic problem is formulated as
\begin{mini!}|s|[2]                   
    {_{\Xi_\text{S}}} 
    { \mathcal{C}_\text{S} = \mathcal{C}_\text{R} + \mathcal{C}_\text{D} + \sum_{s \in \mathcal{S}} \pi_s \,  \mathcal{C}_{\text{B},s} }{\label{prob:ST}}
    {} 
    \addConstraint{\eqref{RM_rup_lim}  \nonumber,}{}{ \quad  \text{Reserve market} }
    \addConstraint{\eqref{DAM_bal} - \eqref{DAM_f} \nonumber,}{}{ \quad  \text{Day-ahead market }}
    \addConstraint{\eqref{BM_bal} - \eqref{BM_f} \nonumber,}{}{ \quad  \text{Balancing market}, \hspace{1.5cm} \forall s \in \mathcal{S}}
\end{mini!}
where $\Xi_\text{S} = \{ \Xi_\text{R} \cup \Xi_\text{D} \cup \Xi_\text{B,s}, \forall s \}$ is the set of decision variables. 
The stochastic co-optimization of energy and reserves attains perfect temporal coordination, as opposed to the sequential model that separates the day-ahead and balancing decisions. Each generator is pre-positioned even out of merit order in a way that allows optimal delivery to the system in case of deviations from the day-ahead schedule. For this reason, we use the stochastic co-optimization approach as a benchmark to our proposed methodology, since it provides a lower bound to the total operational costs.

%% file: 06_MPEC.tex
\section{Reserve and Capacity Allocation Models} \label{sec:RCAM}
This section introduces the concepts and the mathematical formulations that underpin the contributions of this work. The bilevel models in \cite{Dvorkin2019} and \cite{Delikaraoglou2019} are enhanced with a set of upper-level grid partitioning constraints described in \ref{subsec:gp}. These enable the operator to identify a pre-specified number of zones in the system, where zonal reserve markets can be cleared following the problem formulation in \ref{subsec:drp}.The model is complemented with a set of upper-level decision variables that account for the optimal allocation of transmission capacity between energy trading and re-dispatch actions in \ref{subsec:tca}. Lastly, the full problem formulation of the proposed Zonal Preemptive methodology is presented in \ref{subsec:MPECz}. 

\subsection{Grid Partitioning} \label{subsec:gp}
Let $\Theta = (\mathcal{N},\mathcal{L})$ be a directed graph with $\mathcal{N}$ nodes and $\mathcal{L}$ edges describing the single-phase equivalent topology of a power system. The partition of such a graph into $\mathcal{Z}$ connected sub-graphs or zones can be achieved by assigning as many binary variables $x_{n,z} \in \{0,1\}$ as the number of zones to each node. If node $n$ belongs to zone $z$, then $x_{n,z} = 1$; otherwise $x_{n,z} = 0$. Three important properties need to be satisfied in order to get the desired partition: (1) the zones are mutually exclusive; (2) each node belongs to a zone; (3) the sub-graphs determined by the partition are connected, i.e., whichever two points are selected inside a zone, there always exists a path connecting them within the same zone. The first two properties are satisfied with
\begin{equation}
    \sum_{z \in \mathcal{Z}} x_{n,z} = 1, \quad \forall n \in \mathcal{N}, \label{eq:part_xn}  
\end{equation}
whereas, to achieve the third property, this paper adopts the single-commodity flow method presented in \cite{Fan2012}.This method relies on \textit{flowing units}, which bear no physical meaning, but allow to express the connectivity as the ability to reach all nodes in a zone, while staying within its boundaries. This method works by injecting $y_z$ units, i.e., as many as the number of nodes in the $z$-th zone, into a single arbitrary node of each sub-graph and enforcing 
\begin{equation}
    y_z = \sum_{n \in \mathcal{N}} x_{n,z}, \quad \underline{y}_z \leq y_z \leq \overline{y}_z, \hspace{1.5cm} \forall z \in \mathcal{Z}, \label{eq:part_yz}
\end{equation}
where the quantity $y_z$ can be bounded by $\underline{y}_z$ and $\overline{y}_z$ in order to require a minimum or a maximum size of each zone in the system, respectively. A sub-graph is then connected if all the injected units can flow to the nodes in that sub-graph, without violating nodal flow balance and branch flow limit constraints. Nodal flow balance is expressed in a matrix notation as
\begin{equation}
    \boldsymbol{H}_{(\cdot,n)}^\top \boldsymbol{\varphi}_{(\cdot,z)} + c_{n,z} y_z = x_{n,z}, \quad \forall n \in \mathcal{N}, \quad \forall z \in \mathcal{Z}, \label{eq:part_nod_bal}
\end{equation}
where $\boldsymbol{H}_{(\cdot,n)}$ indicates the $n$-th column of the branch incidence matrix, whose $\ell$-th value is 1 if line $\ell$ enters node $n$, -1 if it leaves it, or 0 otherwise and $\boldsymbol{\varphi}_{(\cdot,z)}$ collects the flow of units injected in zone $z$ over all branches in the system. Therefore, the scalar product $\boldsymbol{H}_{(\cdot,n)}^\top \boldsymbol{\varphi}_{(\cdot,z)}$ describes the net in- or out-coming flow of units to or from node $n$ in zone $z$. The bilinear term $ c_{n,z} y_z $ represents instead the injection of $y_z$ flow units into the root nodes defined by $c_{n,z}$, whereas the right-hand-side acts as a sink, i.e., if node $n$ is included in zone $z$, it retains one unit.

Note that unlike \cite{Fan2012}, this novel version of the single-commodity flow method does not require the root nodes $c_{n,z}$ to be pre-specified. This requirement limits the degrees of freedom of the partitioning algorithm, as it relies on the choice of the initial nodes, from which the sub-graphs are generated. This last step is not necessary here, since $c_{n,z}$ is treated as a binary variable, which selects a node where the units are injected. The following constraints are added to ensure that the selected root nodes are mutually exclusive and that only one node per zone is selected as the root, i.e,
\begin{align}
&\sum_{z \in \mathcal{Z}} c_{n,z} \leq 1, && \quad \forall n \in \mathcal{N}, \label{eq:part_cn} \\
&\sum_{n \in \mathcal{N}} c_{n,z} = 1, && \quad \forall z \in \mathcal{Z}. \label{eq:part_cz}
\end{align}
Finally, the branch flow limits are specifically defined to restrict the flow of units $\varphi_{\ell,z}$ to those lines that have both ends included in the same sub-graph. This aspect is modelled using the following constraints
\begin{align}
    & - \Phi_{\text{F}_{\ell,z}} \leq \varphi_{\ell,z} \leq \Phi_{\text{F}_{\ell,z}},  && \quad \forall \ell \in \mathcal{L}, &&& \hspace{-0.5cm}\forall z \in \mathcal{Z},   \label{eq:part_phiF_lim}\\
    & - \Phi_{\text{T}_{\ell,z}} \leq \varphi_{\ell,z} \leq \Phi_{\text{T}_{\ell,z}},  && \quad \forall \ell \in \mathcal{L}, &&& \hspace{-0.5cm}\forall z \in \mathcal{Z},   \label{eq:part_phiT_lim} \\
    & \Phi_{\text{F}_{\ell,z}} = y_z \left(\boldsymbol{H}_{\text{F}_{(\ell,\cdot)}} \boldsymbol{x}_{(\cdot,z)}\right), && \quad \forall \ell \in \mathcal{L}, &&&\hspace{-0.5cm} \forall z \in \mathcal{Z}, \label{eq:part_phiF}  \\
    & \Phi_{\text{T}_{\ell,z}} = y_z \left(\boldsymbol{H}_{\text{T}_{(\ell,\cdot)}} \boldsymbol{x}_{(\cdot,z)}\right), && \quad \forall \ell \in \mathcal{L}, &&& \hspace{-0.5cm}\forall z \in \mathcal{Z}. \label{eq:part_phiT}
\end{align}
where $\boldsymbol{H}_\text{F}$ and $\boldsymbol{H}_\text{T}$ indicate ``from" and ``to" incidence matrices, respectively. For any given sub-graph, the maximum flow of units on each branch is bounded both by $\Phi_{\text{T}_{\ell,z}}$ and $\Phi_{\text{F}_{\ell,z}}$, which are equal to the injected quantity $y_z$, if the line is fully within the sub-graph, or 0, otherwise. This condition is modelled with the scalar products $\boldsymbol{H}_{\text{F}_{(\ell,\cdot)}} \boldsymbol{x}_{(\cdot,z)}$ and $\boldsymbol{H}_{\text{T}_{(\ell,\cdot)}} \boldsymbol{x}_{(\cdot,z)}$, whose values are either 1, if the ``from" or ``to" node of line $\ell$ is included in zone $z$, or 0, if not. Therefore, the flow of units is prevented, unless both scalar products in \eqref{eq:part_phiF}-\eqref{eq:part_phiT} are equal to 1. In this case, $\varphi_{\ell,z}$ is limited by $y_z$, which always represents an upper bound to the highest possible flow of units.  

To summarize, the grid partitioning requires the set of decision variables $\Xi_\text{G} = \{ x_{n,z}, c_{n,z}, \forall n, \forall z; \, y_z, \forall z; \, \varphi_{\ell,z}, \forall \ell, \forall z \}$ constrained by \eqref{eq:part_xn} - \eqref{eq:part_phiT} in the upper-level problem of the proposed bilevel methodology. Section \ref{sec:solution} of the paper describes in detail the linearization of the bilinear terms that appear in constraints \eqref{eq:part_nod_bal}, \eqref{eq:part_phiF} and \eqref{eq:part_phiT} using the big-M approach \cite{Trespalacios2014}.

\subsection{Dynamic Reserve Procurement} \label{subsec:drp}
The proposed methodology allows the grid operator to identify and clear $\mathcal{Z}$ independent reserve markets, each of them corresponding to a zone of the partition. Although the objective remains to minimize the total procurement cost of reserves, zonal requirements can be differentiated while respecting the merit order of generators that participate in each reserve market. The dynamic reserve procurement model constitutes one of the two lower-level problems in the bilevel structure of the proposed methodology and it is formulated as

\begin{mini!}|s|[3]                   
    {_{\Xi_{\text{R}_z}}} 
    { \mathcal{C}_{\text{R},z} = \sum_{g \in \mathcal{G}} \left( C_g^+ r_g^+ + C_g^- r_g^- \right) }{\label{prob:RMz}} 
    {\label{test}} 
    \addConstraint{\sum_{g \in \mathcal{G}} r_{g,z}^+  }{ \geq \lambda_z^+, }{ \quad \forall z \in \mathcal{Z}, \label{RMz_rup_req}}
    \addConstraint{\sum_{g \in \mathcal{G}} r_{g,z}^-  }{\geq \lambda_z^-, }{ \quad \forall z \in \mathcal{Z}, \label{RMz_rdw_req}}
    \addConstraint{0 \leq r_{g,z}^+ }{\leq R_g^+ \left(\boldsymbol{H}_{\text{G}_{(g,\cdot)}} \boldsymbol{x}_{(\cdot,z)}\right), }{ \quad \forall z \in \mathcal{Z},  \quad \forall g \in \mathcal{G}, \label{RMz_rup_lim}}
    \addConstraint{0 \leq r_{g,z}^- }{\leq R_g^- \left(\boldsymbol{H}_{\text{G}_{(g,\cdot)}} \boldsymbol{x}_{(\cdot,z)}\right),  }{ \quad \forall z \in \mathcal{Z}, \quad \forall g \in \mathcal{G}, \label{RMz_rdw_lim}}
    \addConstraint{r_g^+ =}{ \sum_{z \in \mathcal{Z}} r_{g,z}^+,}{ \quad \forall g \in \mathcal{G}, \label{RMz_rup_def}}
    \addConstraint{r_g^- =}{ \sum_{z \in \mathcal{Z}} r_{g,z}^-,}{ \quad \forall g \in \mathcal{G} \label{RMz_rdw_def}}
\end{mini!}
where $\Xi_{\text{R}_z} = \{ r_{g,z}^+, r_{g,z}^-, \forall g, \forall z \}$ is the set of decision variables: $r_{g,z}^+$ and $r_{g,z}^-$ represent up- and downward reserve from generator $g$ in zone $z$, respectively. Note that since $x_{n,z}$ is an upper-level variable, it enters \eqref{prob:RMz} as a parameter, thus rendering the sub-problem a linear one. This structure allows to use the associated Karush-Kuhn-Tucker (KKT) conditions to reformulate the bilevel structure into a mathematical problem with equilibrium constraints (MPEC). The upward and downward zonal requirements $\lambda_z^+$ and $\lambda_z^-$ imposed through 
(14b)-\eqref{RMz_rdw_req} 
also enter this formulation as parameters, since they are upper-level decision variables. Zonal reserve requirements $\lambda_z$ are fulfilled by generators that belong to the corresponding zones. The scalar product in \eqref{RMz_rup_lim}-\eqref{RMz_rdw_lim} between  $\boldsymbol{H}_{\text{G}_{(g,\cdot)}}$ and $\boldsymbol{x}_{(\cdot,z)}$ indicates whether generator $g$ is eligible for providing reserve to zone $z$. Finally, \eqref{RMz_rup_def}-\eqref{RMz_rdw_def} define the overall reserve to be acquired from each generator.

\subsection{Transmission Capacity Allocation} \label{subsec:tca}
\begin{figure}[b]
    \centering
    \includegraphics[width=1\columnwidth]{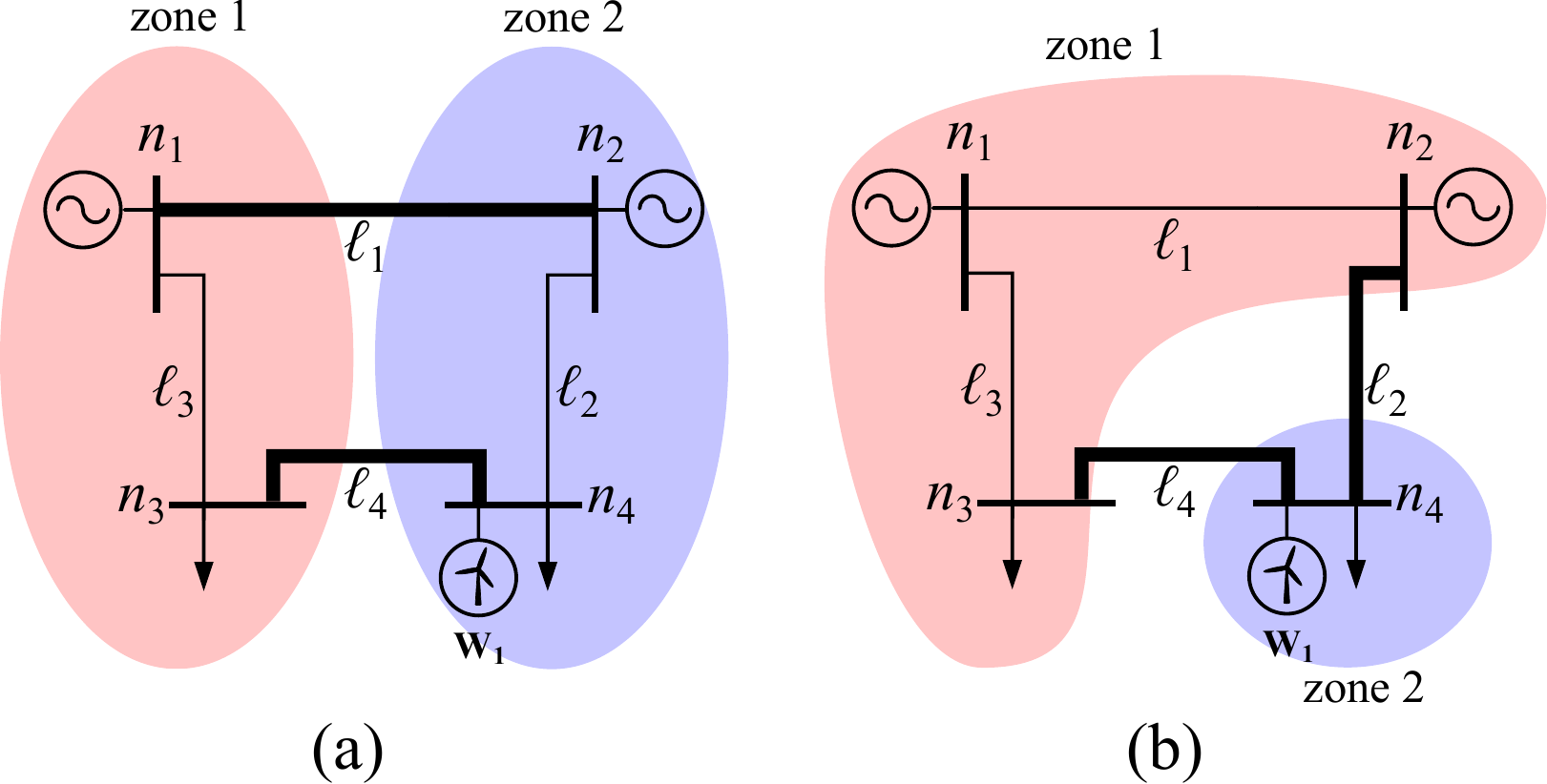}
    \vspace{-0.5cm}
    \caption{ Illustrative case of two possible configurations of grid partitioning on a 4-bus system. Thicker lines indicates cross-zonal interconnections.}
    \label{fig:toymod}
\end{figure}

In cases when flexible resources are concentrated in a certain zone of the system, the grid operator could set aside part of the cross-zonal transmission capacity in order to facilitate the exchange of reserves. This aspect is modelled in the proposed formulation by means of an additional set of upper-level decision variables $\Xi_\text{C} = \{ h_{\ell,z}, \Gamma_{\ell,z}, \Gamma_{\ell} \}$, which defines the available capacity for energy trading on each cross-zonal line. Consider, for example, Fig. \ref{fig:toymod} where a 4-bus system is partitioned in two possible configurations.
Note that as the zones are defined dynamically, so are the cross-zonal lines eligible for reserve exchange, i.e., lines $\ell_1$ and $\ell_4$ in Fig. \ref{fig:toymod}(a) as opposed to lines $\ell_2$ and $\ell_4$ in Fig. \ref{fig:toymod}(b). Therefore, the partitioning identifies endogenously the lines, whose capacity can be set aside for reserve exchange, through an auxiliary integer variable $h_{\ell,z}$, defined as the number of nodes that a line $\ell$ has in zone $z$ according to the following expression
\begin{equation}
     h_{\ell,z} = \boldsymbol{H}_{\text{F}_{(\ell,\cdot)}} \boldsymbol{x}_{(\cdot,z)} + \boldsymbol{H}_{\text{T}_{(\ell,\cdot)}} \boldsymbol{x}_{(\cdot,z)}, \quad \forall \ell \in \mathcal{L}, \, \forall z \in \mathcal{Z}. \label{eq:h} 
\end{equation}
The values that $h_{\ell,z}$ can take are: 0, 1 or 2 and they reflect all possible configurations between lines and zones. In the first case $h_{\ell,z} = 0$, the line is totally outside the considered zone, e.g., line $\ell_2$ with respect to zone 1 in Fig. \ref{fig:toymod}(a); in the second case $h_{\ell,z} = 1$, the line is cross-zonal because only one of the two nodes is included in a zone, e.g., line $\ell_1$ in Fig. \ref{fig:toymod}(a); in the third case, $h_{\ell,z} = 2$ indicates a line that is fully included in the considered zone, e.g., $\ell_1$ in  Fig. \ref{fig:toymod}(b). Only when $h_{\ell,z} = 1$ a portion of the capacity of line $\ell$ is set aside, while the other cases identify domestic lines whose capacity is entirely allocated for energy trading in the day-ahead market. The following set of constraints limits the capacity allocation for reserve exchange $\Gamma_{\ell,z}$ to or from zone $z$ on line $\ell$, 
\begin{align}
    & \Gamma_{\ell,z} \leq \chi F_\ell \, h_{\ell,z}, \quad && \forall \ell \in \mathcal{L}, \quad \forall z \in \mathcal{Z}, \label{eq:gammaR_h} \\
    & \Gamma_{\ell,z} \leq \chi F_\ell \, \left( 2 - h_{\ell,z} \right), \quad && \forall \ell \in \mathcal{L}, \quad \forall z \in \mathcal{Z}, \label{eq:gammaR_2-h} 
\end{align}
where a predefined parameter $\chi$ is included in order to limit the maximum capacity that can be withdrawn from day-ahead market and $F_\ell$ indicates the line rating. When $h_{\ell,z}$ is either 0 or 2, one of the above constraints binds $\Gamma_{\ell,z}$ to be zero, thus preventing any capacity of that line to be set aside. In the remaining case, $h_{\ell,z} = 1$, both \eqref{eq:gammaR_h} and \eqref{eq:gammaR_2-h} state that the share of capacity can be up to the $\chi$\% of the line rating. The remaining constraints include
\begin{align}
    & 0 \leq \Gamma_{\ell,z} \leq \Gamma_\ell, \quad && \forall \ell \in \mathcal{L}, \quad \forall z \in \mathcal{Z}, \label{eq:gamma1} \\
    &  \Gamma_\ell = \frac{1}{2}\sum_z \Gamma_{\ell,z}, \quad && \forall \ell \in \mathcal{L},  \label{eq:gamma2}
\end{align}
that serve a twofold purpose. The first is to enforce non-negativity of $\Gamma_{\ell,z}$, the second is to define $\Gamma_{\ell}$, which is used to define uniquely the value of capacity to be set aside on each line $\ell$, regardless of the zone considered. Note how the use of $\frac{1}{2}$ prevents counting the line capacity twice in \eqref{eq:gamma2}.

Therefore, with transmission capacity allocation day-ahead power flows are bounded by $F_\ell - \Gamma_\ell$, rather than $F_\ell$. This limits the expected power flows at the day-ahead stage, in order to ensure that enough transmission capacity is available during real-time operation.

\subsection{Zonal Preemptive Problem Formulation} \label{subsec:MPECz}
The proposed methodology builds upon recent work that adopted a stochastic bilevel framework for setting reserve requirements \cite{Dvorkin2019}-\cite{Delikaraoglou2019}. While previous studies considered either a single zone or a predefined zonal setup, we improve the positioning and accessibility of reserves by defining zone boundaries together with their reserve requirements. The complete problem formulation, where both reserve and transmission capacity are dynamically allocated, is formulated as  

\begin{mini!}|s|[2]                   
    {_{\Xi_{\text{M}_z}}} 
    { \mathcal{C}_{\text{M}_z} = \mathcal{C}_\text{R} + \mathcal{C}_\text{D} + \sum_{s \in \mathcal{S}} \pi_s \,  \mathcal{C}_{\text{B},s} }{\label{prob:MPECz}}
    {} 
    \addConstraint{ (r_g^+, r_g^-) \in  \text{arg} \left\{
        \begin{aligned}
            & \underset{\textstyle \Xi_{\text{R}_z}'}{\text{minimize}} \quad \mathcal{C}_{R_z} \\
            & \text{subject to} \\
            & \text{constraints } (14\text{b}) - \eqref{RMz_rdw_def} 
        \end{aligned} \right\}, \label{lowlev:RMz}  }
    \addConstraint{ (p_g, w_j) \in  \text{arg} \left\{
        \begin{aligned}
            & \underset{\textstyle \Xi'_D}{\text{minimize}} \quad \mathcal{C}_\text{D} \\
            & \text{subject to} \\
            & \text{constraints } \eqref{DAM_bal} - \eqref{DAM_f_lim} %
        \end{aligned} \right\}, \label{lowlev:DAM} }
    \addConstraint{ \lambda^+_z \geq 0, \quad \lambda^-_z \geq 0,}{}{ \hspace{-1cm} \forall z \in \mathcal{Z}, \label{con:ResReq}}
    \addConstraint{\eqref{BM_bal} - \eqref{BM_f} \nonumber, \quad \text{Balancing market},}{}{\hspace{-1cm} \forall s \in \mathcal{S},}
    \addConstraint{\eqref{eq:part_xn} - \eqref{eq:part_phiT} \nonumber, \hspace{0.50cm}  \text{Grid partitioning}, }
    \addConstraint{\eqref{eq:h} - \eqref{eq:gamma2} \nonumber, \quad \text{Capacity allocation},}{}{\hspace{0.5cm} }
\end{mini!}
where $\Xi_{\text{M}_z} = \{ \lambda_z^+, \lambda_z^-, \forall z \cup \Xi_\text{R} \cup \Xi_\text{D} \cup \Xi_\text{B,s}, \forall s \cup \Xi_\text{G} \cup \Xi_\text{C} \}$ is the set of upper-level decision variables. This comprises: zonal reserve requirements $\lambda_z$; reserve, day-ahead and balancing market decision variables, which are constrained by the corresponding lower level problems; grid partitioning and transmission capacity allocation variables $\Xi_\text{G}$ and $\Xi_\text{C}$, respectively. Lower-level problem \eqref{lowlev:RMz} accounts for the dynamic reserve allocation strategy described in \ref{subsec:drp}, whereas \eqref{lowlev:DAM} is the same day-ahead market clearing model as in model \eqref{prob:DAM} where line ratings $F_\ell$ are substituted with $(F_\ell - \Gamma_\ell)$.

The solution of \eqref{prob:MPECz} provides the grid operator with a suggestion of how to split the system into a pre-defined number of zones, where reserve markets could be cleared independently of one another, thus remaining fully compatible with the least-cost merit-order principle in each zonal reserve market. Therefore, unlike the preemptive model in \cite{Dvorkin2019}, it additionally allows to identify those portions of the grid where reserve is required the most and it sets different reserve requirements $\lambda_z$, accordingly. This latter property resembles the ability of the purely stochastic model to preposition reserves anywhere in the system down to a generator-specific resolution. However, since such a degree of freedom cannot be attained in practice, the proposed formulation circumvents this aspect by enforcing a minimal zonal size.

%% file: 07_Solution.tex
\section{Solution Approach} \label{sec:solution}
All bilinear terms that appear in the grid partitioning constraints can be expressed as a product between a binary and an integer variable, i.e., $y_z \left(\boldsymbol{H}_{\text{F}_{(\ell,\cdot)}} \boldsymbol{x}_{(\cdot,z)}\right)$, $y_z \left(\boldsymbol{H}_{\text{T}_{(\ell,\cdot)}} \boldsymbol{x}_{(\cdot,z)}\right)$ and $c_{n,z} y_z$. The linearization of these terms is illustrated for the latter case, by introducing an auxiliary variable $u_{n,z}$ that replaces the product $c_{n,z} y_z$ in \eqref{eq:part_nod_bal} according to the Big-M approach \cite{Trespalacios2014}. The following constraints are added
\begin{align}
    y_z - M \left( 1 - c_{n,z} \right) &\leq u_{n,z} \leq y_z - m \left( 1 - c_{n,z} \right) \\
    m \cdot c_{n,z} &\leq u_{n,z} \leq  M \cdot c_{n,z} 
\end{align}
where $m = 1$ and $M = \mathcal{N}$, i.e., the number of nodes in the system. Note that the specific values of $m$ and $M$ are used for all the bilinear terms that appear in the grid partitioning constraints. These values are straightforward to derive: each bilinear term is either 0 or equal to the sub-graph cardinality $y_z$, thus 1 and $N$ always represent valid bounds.   

For any feasible partition of the system defined by the upper level variables, each lower level problem is linear and convex. Thus, the bilevel problem is reformulated as an MPEC, where each lower-level problem is replaced by the corresponding KKT conditions. This step introduces additional auxiliary binary variables in order to linearize the complementarity slackness constraints in the KKT conditions. The MPEC problem is then recast as a single-level mixed-integer linear problem (MILP) by using the Big-M method. This solution approach is typically used in power systems research, although authors of \cite{Pineda2019} recently pointed out some critical limitations. Finally, considering the special structure of the problem, whose second-stage constraints are independent per scenario, a multi-cut Bender's decomposition scheme is implemented \cite{Conejo2006}. The complete set of KKT conditions of the lower-level problems, along with the formulations of the Bender's master problem and sub-problems, are provided in the electronic companion of the paper.

%% file: 08_CaseStudy.tex
\section{Results} \label{sec:results}

\subsection{Wind Power Scenarios}
In this paper, probabilistic forecast errors of wind power generation are assumed to follow a Beta distribution, whose parameters are calculated according to \cite{Bludszuweit2008} and the error variance follows a quadratic function of the \textit{per unit} point forecast. The spatial correlation structure in wind power generation at different locations is modelled by means of a Gaussian copula function with a rank correlation matrix based on actual wind power realizations from the Danish system \cite{Papaefthymiou2009}. A large number of scenarios is then generated by sampling the resulting multivariate joint probabilistic forecast for a single time-period. In order to keep computational tractability in the stochastic programs, scenario sets $\Omega_i$ are reduced accordingly to 100 realizations using the fast-forward scenario reduction technique \cite{Dupacova2000}. 

\subsection{Stability Analysis (IEEE RTS-24 System)}
The proposed methodology is showcased on a modified version of the IEEE RTS-24 system, whose detailed parameters are available in \cite{rts24}. In particular, three lines are de-rated and six wind farms with 200 MV installed capacity are included in the system at selected locations. Simulation results are benchmarked against the stochastic co-optimization of energy and reserve \eqref{prob:ST} and the conventional approach of sequentially cleared markets, i.e.,  \eqref{prob:RM}, \eqref{prob:DAM} and \eqref{prob:BM}. Reserve requirements $\Lambda$ in model \eqref{prob:RM} are calculated as
\begin{align}
    \Lambda^+ & = \widehat{W}_\text{tot} - \widehat{\boldsymbol{F}}^{-1}_{W}(q)   \label{eq:Wmin}\\
    \Lambda^- & = \widehat{\boldsymbol{F}}^{-1}_{W}(1-q) - \widehat{W}_\text{tot}; \label{eq:Wmax}  
\end{align}
where $\widehat{W}_\text{tot}$ and $\widehat{\boldsymbol{F}}_W$ represent the expected value and the predictive CDF of the total wind power probabilistic forecast, respectively, while the pre-determined quantile $q$ of the distribution is chosen in line with grid operator's risk aversion. The proposed methodology is tested either with or without transmission capacity allocation, where three values of the parameter $\chi$ are considered in the former case and a minimal zonal size of 4 nodes is always required. 

\begin{figure}[t]
    \centering
    \includegraphics[width=1\columnwidth]{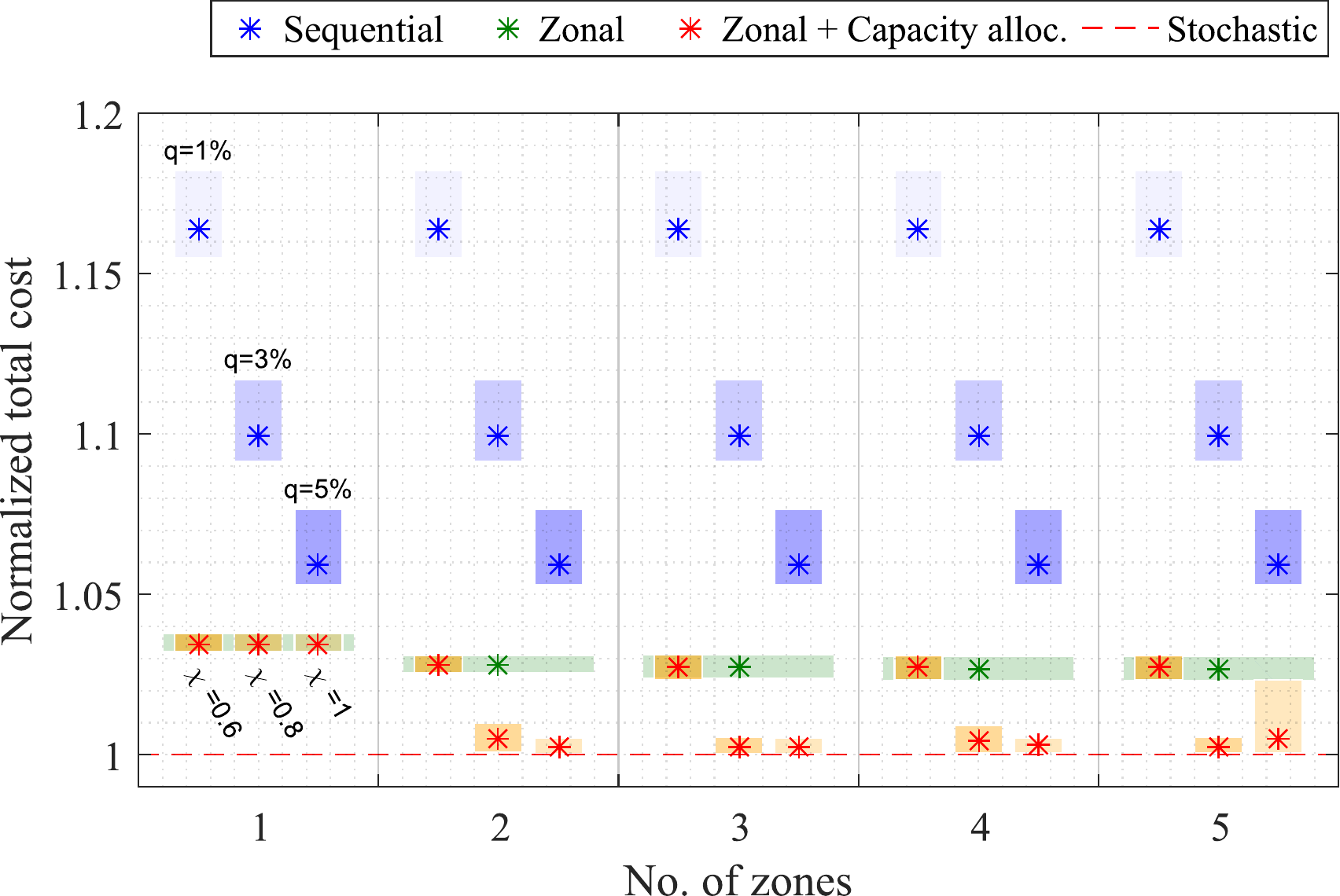}
    \vspace{-0.5cm}
    \caption{ Stability analysis in the RTS 24 bus system. Asterisks represent the mean values, upper and lower edges of rectangles represent max and min values, respectively.}
    \label{fig:robust_all}
    \vspace{-0.2cm}
\end{figure}

In order to test the stability of the considered models against small deviations in the uncertain wind power generation, up- and downward reserve levels obtained with scenario set $\Omega_1$ are plugged into \eqref{prob:DAM} and \eqref{prob:BM}, where the uncertainty is described by 10 different scenario sets $\Omega_i$ ($i=2,\dots,11$), based on the same multivariate probabilistic forecast. Figure \ref{fig:robust_all} shows the corresponding total cost, which are normalized with the solution of the stochastic model obtained with each set $\Omega_i$. It stands out that the zonal approach outperforms the sequential one both in terms of stability and cost effectiveness, regardless of the chosen quantile $q$ for setting reserve levels. The zonal model shows an improvement with just 2 zones, whereas it coincides with model \cite{Dvorkin2019} when a single zone is considered. 

The effect of allowing transmission capacity allocation on cross-zonal lines is to lower the costs further, provided that more than 60\% of eligible line capacities is withdrawn from the day-ahead market. This result resembles the line-switching approach, where the cost effectiveness of a dispatch can be improved if power flows are re-routed by switching off selected lines. In this case, reducing the capacity allocated to the day-ahead market ensures that enough headroom is available for balancing the system, thus avoiding bottlenecks that would result in expensive corrective actions. 

\subsection{Cost Breakdown}

\begin{table}[t]
\centering
\caption{Cost breakdown of selected models}
\label{tab:stagecost}
\begin{tabular}{cccccc}
\toprule
\multicolumn{1}{l}{} &  & \textbf{\begin{tabular}[c]{@{}c@{}}Reserve \\ cost \\ {[}k\${]}\end{tabular}} & \textbf{\begin{tabular}[c]{@{}c@{}}Day-ahead\\ cost\\ {[}k\${]}\end{tabular}} & \textbf{\begin{tabular}[c]{@{}c@{}}Balancing\\ cost \\ {[}k\${]}\end{tabular}} & \textbf{\begin{tabular}[c]{@{}c@{}}Total\\ cost\\ {[}k\${]}\end{tabular}} \\ \cmidrule(l){3-6} 
\multicolumn{2}{c}{\textbf{Sequential}} & 6.015 & 24.41 & 0.322 & 30.75 \\
\multicolumn{2}{c}{\textbf{Stochastic}} & 3.135 & 25.53 & -0.960 & 27.70 \\ \cmidrule(l){3-6} 
\multicolumn{1}{l}{} & \multicolumn{1}{l}{} & \multicolumn{4}{c}{\textbf{\begin{tabular}[c]{@{}c@{}}No capacity allocation \\ $\chi$ = 0\%\end{tabular}}} \\ \cmidrule(l){3-6} 
\multirow{4}{*}{\rotatebox[origin=c]{90}{\textbf{Zonal}}} & \textbf{Z = 1} & 3.810 & 24.34 & 0.432 & 28.59 \\
 & \textbf{Z = 2} & 3.861 & 24.09 & 0.455 & 28.40 \\
 & \textbf{Z = 3} & 3.903 & 24.02 & 0.440 & 28.36 \\
 & \textbf{Z = 4} & 3.901 & 24.01 & 0.434 & 28.35 \\ \cmidrule(l){3-6} 
\multicolumn{1}{l}{} & \multicolumn{1}{l}{} & \multicolumn{4}{c}{\textbf{\begin{tabular}[c]{@{}c@{}}Capacity allocation \\ $\chi$ = 100\%\end{tabular}}} \\ \cmidrule(l){3-6} 
\multirow{4}{*}{\rotatebox[origin=c]{90}{\textbf{Zonal}}} & \textbf{Z = 1} & 3.810 & 24.34 & 0.432 & 28.59 \\
 & \textbf{Z = 2} & 3.160 & 25.54 & -0.997 & 27.70 \\
 & \textbf{Z = 3} & 3.216 & 25.48 & -0.990 & 27.70 \\
 & \textbf{Z = 4} & 3.169 & 25.54 & -1.003 & 27.70 \\ \bottomrule
\end{tabular}
\end{table}

Table \ref{tab:stagecost} shows the cost breakdown of selected models solved with the same scenario set $\Omega_1$. The conventional model in this case relies on the top and bottom 3\% of the total wind power distribution for setting reserve requirements. This approach results in higher cost for reserves, as it cannot account neither for their location in the system nor for the network constraints that might limit their accessibility. Instead, the preemptive model with a single zone is able to regulate reserve requirements based on expected re-dispatch actions. Although improving the results considerably, this approach still relies on a single reserve market and thus it follows the merit order of generators' reserve capacity offers. The implication is that while total reserve levels can be fine-tuned, their location and position in the system cannot. 

\begin{figure}[t]
    \centering
    \includegraphics[width=1\columnwidth]{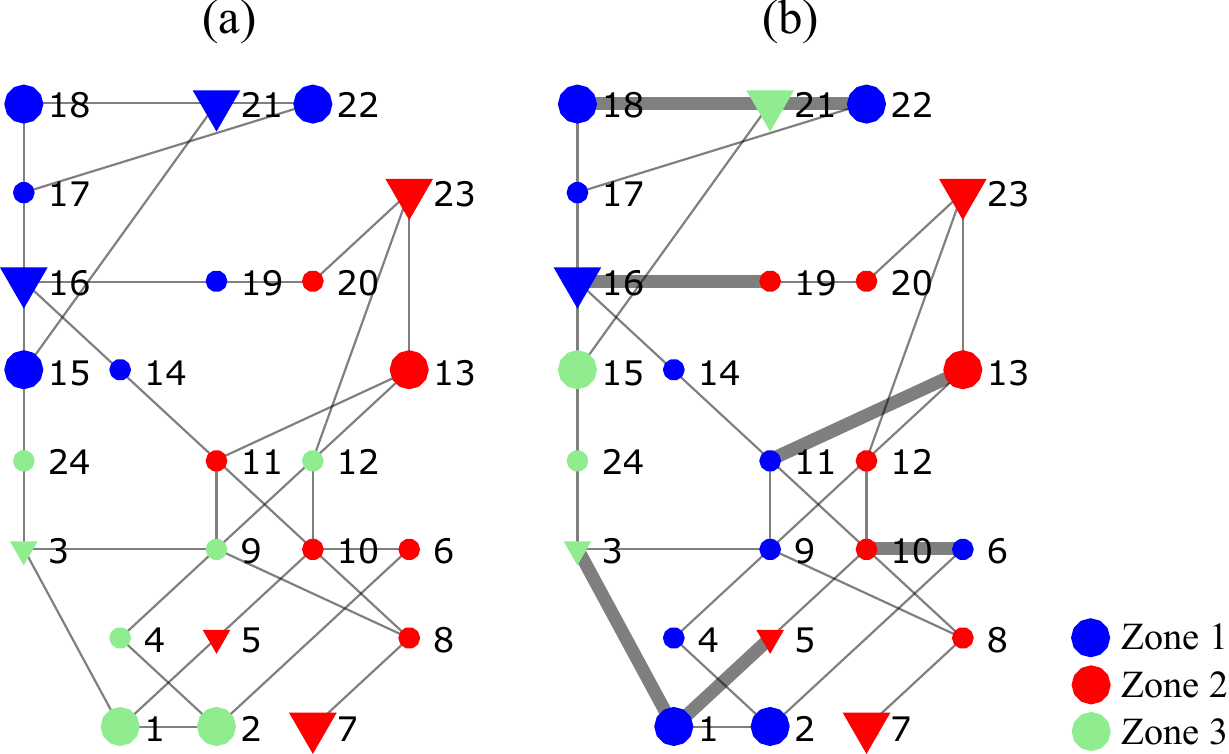}
    \vspace{-0.5cm}
    \caption{ Partition of the RTS-24 system into 3 zones without (a) and  with (b) transmission capacity allocation on selected lines. Triangles indicate wind power, large markers indicate the presence of generators.}
    \label{fig:partitions}
    \vspace{-0.2cm}
\end{figure}

This aspect motivates the introduction of a zonal setup that provides the grid operator with additional flexibility to approximate the ideal solution. As the number of zones increases, so does the ability to lower the costs towards the stochastic model and to optimally allocate reserves. Note that the partitioning in the proposed methodology is solely driven by the total expected costs and it does not require any root node to be pre-specified, which could introduce a degree of arbitrariness in the partition. Therefore, it inherently considers the availability of reserves in the system, their procurement and their activation costs given the network limitations. Table \ref{tab:zoncost} summarizes the zonal reserve costs referring to the case of 3 zones in Fig. \ref{fig:partitions}. The zonal setup allows to procure nearly 60\% of total requirements from zone 2, where the cost per MW is lower. Zone 3 instead procures less reserve, but from more expensive generators ensuring that enough balancing power is located close to wind farms at nodes 3 and 5. 

Two effects are evident as we include transmission capacity allocation on cross-zonal lines: the first is that reserve costs decrease, since more power can be reserved from cheaper generators, while ensuring that they can deliver it to the grid; the second is that day-ahead costs increase, as we reduce the network capacity available for energy trading. A consequence of this latter aspect is that less wind power will be dispatched at this stage. To compensate for that and to avoid expensive wind curtailment penalties, more downward reserve needs to be procured and activated, as can be seen from Table \ref{tab:zoncost}.

\subsection{RTS-96 Case Study}
\begin{table}[t]
\centering
\caption{Zonal reserve cost of grid partitions in Fig. \ref{fig:partitions}}
\label{tab:zoncost}
\begin{tabular}{@{}>{\raggedleft}p{0.85cm}>{\centering}p{0.45cm}>{\centering}p{0.45cm}>{\centering}p{0.45cm}>{\centering}p{0.93cm}>{\centering}p{0.45cm}>{\centering}p{0.45cm}>{\centering}p{0.45cm}>{\centering\arraybackslash}p{0.93cm}@{}}
\toprule
 & \multicolumn{4}{c}{\textbf{\begin{tabular}[c]{@{}c@{}}(a) No capacity allocation \\ $\chi$ = 0\%\end{tabular}}} & \multicolumn{4}{c}{\textbf{\begin{tabular}[c]{@{}c@{}}(b) Capacity allocation \\ $\chi$ = 100\%\end{tabular}}} \\ \cmidrule(l){2-9} 
 & \multicolumn{3}{c}{\begin{tabular}[c]{@{}c@{}}Reserve volume \\ {[}MW{]}\end{tabular}} & \begin{tabular}[c]{@{}c@{}}Avg.\\ cost \\ {[}\$/MW{]}\end{tabular} & \multicolumn{3}{c}{\begin{tabular}[c]{@{}c@{}}Reserve volume  \\ {[}MW{]}\end{tabular}} & \begin{tabular}[c]{@{}c@{}}Avg. \\ cost \\ {[}\$/MW{]}\end{tabular} \\ \cmidrule(lr){2-4} \cmidrule(lr){6-8}
 & \textit{Up} & \textit{Dw} & \textit{Total} &  & \textit{Up} & \textit{Dw} & \textit{Total} &  \\ \midrule
\textbf{Zone 1} & 60.0 & 35.7 & 95.7 & 15 & 0 & 67.9 & 67.9 & 8.54 \\
\textbf{Zone 2} & 75.9 & 112.1 & 188.0 & 12 & 27.5 & 143.9 & 171.4 & 9.72 \\
\textbf{Zone 3} & 40.0 & 0 & 40.0 & 14.3 & 30.0 & 30.0 & 60.0 & 13.33 \\ \midrule
\textbf{Total} & 175.9 & 147.8 & 323.7 & 12.7 & 57.5 & 241.8 & 299.3 & 10.23 \\ \bottomrule
\end{tabular}
\end{table}

The proposed methodology is also tested in the 3-area RTS-96 system, for which relevant data is taken from \cite{Dvorkin2019}. The system is considered during the peak hour with a total demand of 7.5 GW, 18\% of which is covered by wind power located in 5 locations. Original line ratings are used, whereas the minimum power output of controllable units is set to 0. In order to exclude the generator-specific resolution of the stochastic approach, a minimal zonal size of 10 nodes has been enforced in this system.

\begin{figure}[t]
    \centering
    \includegraphics[width=1\columnwidth]{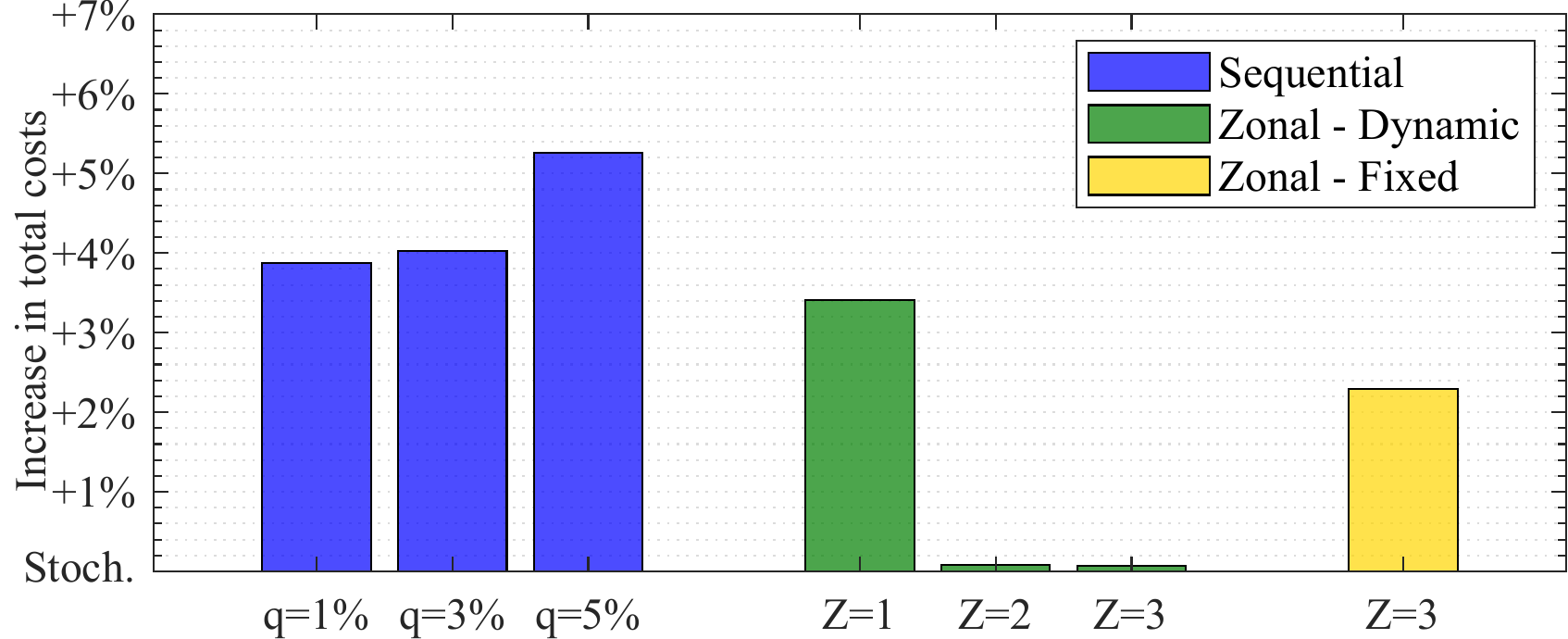}
    \vspace{-0.5cm}
    \caption{Increase in total expected costs for the load demand peak hour in the RTS-96 system. Costs normalized with the stochastic solution. }
    \label{fig:rts96costs}
\end{figure}

\begin{figure}[t]
    \centering
    \includegraphics[width=1\columnwidth]{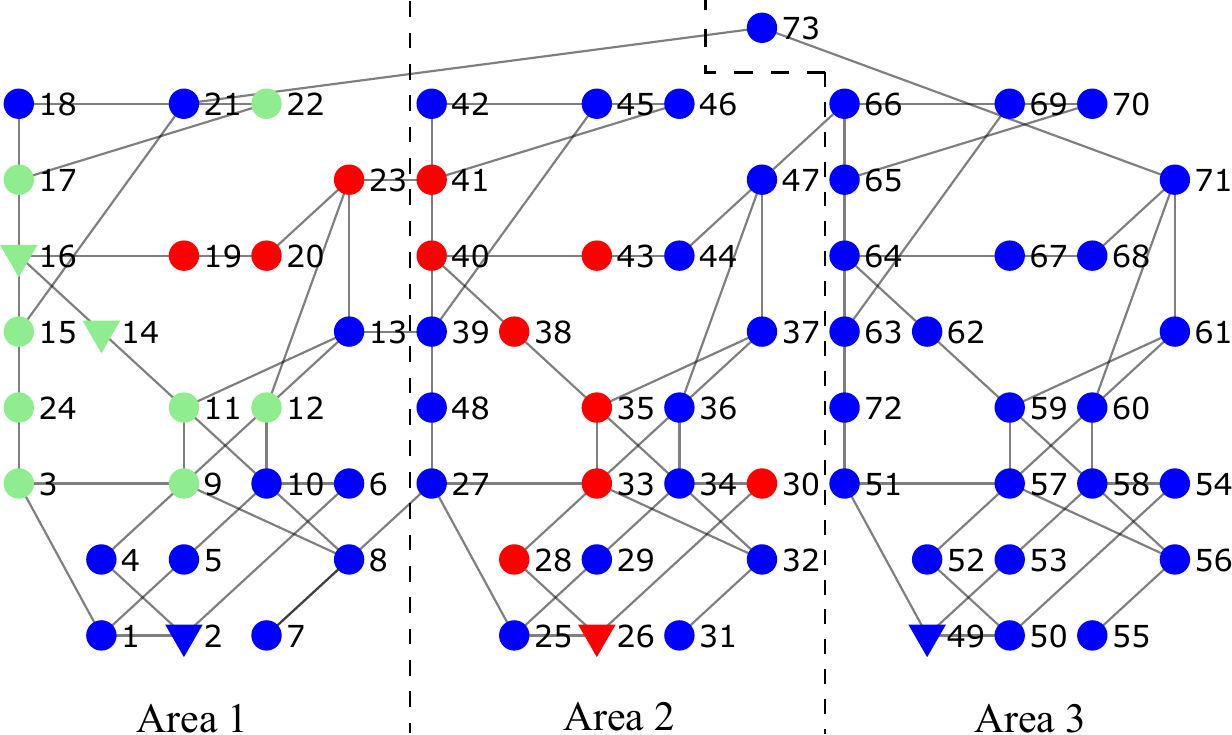}
    \vspace{-0.5cm}
    \caption{Partition of the 3-area IEEE RTS-96 system according to the Zonal Preemptive model with $\chi = 0$. Triangles indicate wind farms.}
    \label{fig:rts96areas}
    \vspace{-0.2cm}
\end{figure}

Figure \ref{fig:rts96costs} shows the resulting total expected costs, where we compare the proposed methodology to the sequential approach with varying reserve requirements (i.e. corresponding to different quantiles $q$ in \eqref{eq:Wmax} and \eqref{eq:Wmin}) and a zonal model with a predetermined partitioning variable $x_{n,z}$, according to the standard partitioning of this system into three zones. In this way, we isolate the contribution of the flexible zone boundaries definition provided by the proposed Zonal Preemptive approach. The partition into 3 zones is shown in Fig. \eqref{fig:rts96areas} together with the common subdivision of the RTS-96 system into 3 areas. The total expected costs indicate that tuning reserve requirements while considering a single reserve market, i.e., $Z=1$, does not result in significant savings, as opposed to the sequential approach. The Zonal Preemptive approach with a fixed partition that adheres to the 3 areas in Fig. \ref{fig:rts96areas} performs better than the single zone. However, as the zones are dynamically determined by the partitioning variables $x_{n,z}$, the total costs fall near the lower bound represented by the stochastic co-optimization of energy and reserve. It suffices to split the system into 2 zones to stay within the 0.1\% increase from the lower bound, even without allocating transmission capacity on cross-zonal lines. 

The resulting large-scale MILP problem is solved with Gurobi setting a 0.1\% optimality gap on a quad-core laptop with 8 GB of RAM and 2.4 GHz of CPU. Bender's decomposition converged in $3.50 \cdot10^2$ s with 1 zone, $2.15\cdot10^3$ s with 2 zones and $4.04\cdot10^4$ s with 3 zones. 


%% file: 09_Conclusion.tex
\section{Conclusion} \label{sec:conclusion}
This paper described a novel methodology for reserve procurement that further approximates the efficiency of the stochastic co-optimization of energy and reserves in terms of total operating costs, while still respecting the existing market rules. Building upon recent work on stochastic bilevel optimization, we embed grid partitioning constraints in the upper-level problem and use them to determine not only the zonal reserve requirements but the zonal boundaries as well. 

Unlike other partitioning schemes, our methodology is solely driven by the total expected system costs and the most recent uncertainty forecasts, instead of relying on historical data that may not reflect the actual system state. The proposed model allows grid operators to perform a dynamic zoning of the system for reserve procurement, depending upon generation uncertainty and network limitations. In addition, this zonal preemptive model can contribute to the ongoing policy discussion towards a common European reserve capacity market, where reserve zones are dynamically defined upon system conditions instead of geographical borders. Simulation results show that the stochastic lower bound can be adequately approximated with only two zones, even if a minimal zonal size is required. This result suggests that the computational burden of the proposed approach can be reduced by limiting the number of reserve zones, without a major efficiency loss in terms of expected system cost. Moreover, the combination of dynamic reserve procurement with cross-zonal transmission capacity allocation has shown to be beneficial in highly congested system. Setting aside part of the available transmission capacity grants the grid operator additional flexibility to approach the efficiency of the stochastic dispatch. 

Future work will address the current limitations of the proposed methodology considering inter-temporal constraints, which may affect the partitioning of the system and the deployment of reserves.

%% file: 97_Appendix.tex

%

\section*{Companion Paper}
This document serves as the electronic companion of paper "Dynamic Reserve and Transmission Capacity Allocation in Wind-Dominated Power Systems". Appendix \ref{app:A} presents the Karush-Kuhn-Tucker (KKT) conditions of the dynamic reserve procurement problem in Section \ref{subsec:drp} and the day-ahead market clearing in Section \ref{sec:markets}. Appendix \ref{app:B} presents the multi-cut Bender's decomposition scheme that is used to solve the proposed methodology on large-scale systems. 

\appendices
\section{}\label{app:A}
The dual variables in the KKT conditions are indicated as $\gamma^{*}$ for constraint $(*)$, where equation numbers are referred to the main paper. The corresponding optimization problems are repeated in a standard notation to ease the identification of dual variables. Symbol $\bot$ indicates the complementarity conditions between the constraints and the rest of the symbols is in accordance with the nomenclature in the main paper. 

\subsection{Zonal reserve market} 
\subsubsection{Problem formulation}
\begin{mini!}|s|[3]                   
    {_{\Xi_{\text{R}_z}}} 
    { \mathcal{C}_{\text{R},z} = \sum_{g \in \mathcal{G}} \left( C_g^+ r_g^+ + C_g^- r_g^- \right) }{\label{prob:kktRMz}} 
    {} 
    \addConstraint{ }{ \lambda_z^+ - \sum_{g \in \mathcal{G}} r_{g,z}^+ \leq 0, }{ \quad \forall z \in \mathcal{Z}, \label{kktRMz_rup_req}}
    \addConstraint{ }{ \lambda_z^- - \sum_{g \in \mathcal{G}} r_{g,z}^-  \leq 0, }{ \quad \forall z \in \mathcal{Z}, \label{kktRMz_rdw_req}}
    \addConstraint{ r_{g,z}^+ }{- R_g^+ \left(\boldsymbol{H}_{\text{G}_{(g,\cdot)}} \boldsymbol{x}_{(\cdot,z)}\right) \leq 0, }{ \quad \forall z \in \mathcal{Z},  \quad \forall g \in \mathcal{G}, \label{kktRMz_rup_lim1}}
    \addConstraint{ r_{g,z}^- }{- R_g^- \left(\boldsymbol{H}_{\text{G}_{(g,\cdot)}} \boldsymbol{x}_{(\cdot,z)}\right) \leq 0,  }{ \quad \forall z \in \mathcal{Z}, \quad \forall g \in \mathcal{G}, \label{kktRMz_rdw_lim1}}
    \addConstraint{ -r_{g,z}^+ \leq 0 }{}{\quad \forall z \in \mathcal{Z}, \quad \forall g \in \mathcal{G}, \label{kktRMz_rup_lim2}}
    \addConstraint{ -r_{g,z}^- \leq 0 }{}{\quad \forall z \in \mathcal{Z}, \quad \forall g \in \mathcal{G}, \label{kktRMz_rdw_lim2}}
    \addConstraint{r_g^+ -}{ \sum_{z \in \mathcal{Z}} r_{g,z}^+ = 0,}{ \quad \forall g \in \mathcal{G}, \label{kktRMz_rup_def}}
    \addConstraint{r_g^- -}{ \sum_{z \in \mathcal{Z}} r_{g,z}^- = 0,}{ \quad \forall g \in \mathcal{G} \label{kktRMz_rdw_def}}
\end{mini!}

\subsubsection{KKT conditions} 
\begin{align}
    & C_g^+ - \gamma^{(25\text{b})}_z + \gamma^{\eqref{kktRMz_rup_lim1}}_{g,z} - \gamma^{\eqref{kktRMz_rup_lim2}}_{g,z} + \gamma^{\eqref{kktRMz_rup_def}}_g = 0, &&\, \forall g, \, \forall z, \label{KKT_RMz_1}\\ 
    & C_g^- - \gamma^{\eqref{kktRMz_rdw_req}}_z + \gamma^{\eqref{kktRMz_rdw_lim1}}_{g,z} - \gamma^{\eqref{kktRMz_rdw_lim2}}_{g,z} + \gamma^{\eqref{kktRMz_rdw_def}}_g = 0, &&\, \forall g, \, \forall z, \\
    & 0 \geq \lambda_z^+ - \sum_{g \in \mathcal{G}} r_{g,z}^+ \quad \bot \quad \gamma^{(25\text{b})}_z \geq 0, &&\, \forall z \\
    & 0 \geq \lambda_z^- - \sum_{g \in \mathcal{G}} r_{g,z}^- \quad \bot \quad \gamma^{\eqref{kktRMz_rdw_req}}_z \geq 0, &&\, \forall z \\
    & 0 \geq r_{g,z}^+ - R_g^+ \left(\boldsymbol{H}_{\text{G}_{(g,\cdot)}} \boldsymbol{x}_{(\cdot,z)}\right) \, \bot \, \gamma^{\eqref{kktRMz_rup_lim1}}_{g,z} \geq 0, &&\, \forall g, \forall z \\
    & 0 \geq r_{g,z}^- - R_g^- \left(\boldsymbol{H}_{\text{G}_{(g,\cdot)}} \boldsymbol{x}_{(\cdot,z)}\right) \, \bot \, \gamma^{\eqref{kktRMz_rdw_lim1}}_{g,z} \geq 0, &&\, \forall g, \forall z \\
    & 0 \geq -r_{g,z}^+ \quad \bot \quad \gamma^{\eqref{kktRMz_rup_lim2}}_{g,z} \geq 0, && \forall g, \forall z \\
    & 0 \geq -r_{g,z}^- \quad \bot \quad \gamma^{\eqref{kktRMz_rdw_lim2}}_{g,z} \geq 0, && \forall g, \forall z \\
    & r_g^+ - \sum_{z \in \mathcal{Z}} r_{g,z}^+ = 0, &&\forall g, \\
    & r_g^- - \sum_{z \in \mathcal{Z}} r_{g,z}^- = 0, &&\forall g. \label{KKT_RMz_last}
\end{align}

\newpage

\subsection{Day-ahead market}
\subsubsection{Problem formulation}
\begin{mini!}|s|[3]                   
    {_{\Xi_\text{D}}} 
    { \mathcal{C}_\text{D} = \sum_{g \in \mathcal{G}} C_g p_g }{\label{prob:kktDAM}} 
    {} 
    \addConstraint{\sum_{g \in \mathcal{G}} p_g + \sum_{j \in \mathcal{J}} w_j - \sum_{n \in \mathcal{N}} D_n }{= 0 , }{\label{kktDAM_bal}}    
    \addConstraint{\underline{P}_g + r_g^{-,*} - p_g \leq 0 }{}{ \hspace{0.2cm} \forall g \in \mathcal{G}, \label{kktDAM_p_lim1}}
    \addConstraint{  p_g - \overline{P}_g + r_g^{+,*} \leq 0, }{}{ \hspace{0.2cm} \forall g \in \mathcal{G}, \label{kktDAM_p_lim2}}
    \addConstraint{ - w_j \leq 0, }{}{ \hspace{0.2cm} \forall j \in \mathcal{J}, \label{kktDAM_w_lim1}}
    \addConstraint{ w_j - \widehat{W}_j \leq 0, }{}{ \hspace{0.2cm} \forall j \in \mathcal{J}, \label{kktDAM_w_lim2}}
    \addConstraint{}{ \boldsymbol{M}_{(\ell,\cdot)} \left( \boldsymbol{H}_\text{G}^\top \boldsymbol{p} + \boldsymbol{H}_\text{J}^\top \boldsymbol{w} - \boldsymbol{D} \right)  - F_\ell \leq 0, }{\hspace{0.2cm} \forall \ell \in \mathcal{L}, \label{kktDAM_f_lim1}}
    \addConstraint{  }{ - F_\ell -\boldsymbol{M}_{(\ell,\cdot)} \left( \boldsymbol{H}_\text{G}^\top \boldsymbol{p} + \boldsymbol{H}_\text{J}^\top \boldsymbol{w} - \boldsymbol{D} \right) \leq 0,}{\hspace{0.2cm} \forall \ell \in \mathcal{L}, \label{kktDAM_f_lim2}}
\end{mini!}

\subsubsection{KKT conditions}
\begin{align}
\begin{split}
     & C_g + \gamma^{(36\text{b})} - \gamma^{\eqref{kktDAM_p_lim1}}_g +  \gamma^{\eqref{kktDAM_p_lim2}}_g \ldots \label{KKT_DA_1} \\ 
     & + \sum_{\ell \in \mathcal{L}} ( \gamma^{\eqref{kktDAM_f_lim1}}_\ell - \gamma^{\eqref{kktDAM_f_lim2}}_\ell ) \boldsymbol{M}_{(\ell,\cdot)} \boldsymbol{H}_\text{G}^\top \mathds{1}_g = 0,  
\end{split} && \quad \forall g \in \mathcal{G}, \\
\begin{split}
     & \gamma^{(36\text{b})} - \gamma^{\eqref{kktDAM_w_lim1}} + \gamma^{\eqref{kktDAM_w_lim2}} \ldots \\ 
     & + \sum_{\ell \in \mathcal{L}} ( \gamma^{\eqref{kktDAM_f_lim1}}_\ell - \gamma^{\eqref{kktDAM_f_lim2}}_\ell ) \boldsymbol{M}_{(\ell,\cdot)} \boldsymbol{H}_\text{J}^\top \mathds{1}_j = 0,     
\end{split} && \quad \forall j \in \mathcal{J}, \\
& \sum_{g \in \mathcal{G}} p_g + \sum_{j \in \mathcal{J}} w_j - \sum_{n \in \mathcal{N}} D_n = 0 \\
& 0 \geq \underline{P}_g + r_g^{-,*} - p_g  \quad \bot \quad \gamma^{\eqref{kktDAM_p_lim1}}_g \geq 0, && \quad \forall g \in \mathcal{G}, \\
& 0 \geq  p_g - \overline{P}_g + r_g^{+,*}   \quad \bot \quad \gamma^{\eqref{kktDAM_p_lim2}}_g \geq 0, && \quad \forall g \in \mathcal{G}, \\
& 0 \geq - w_j \quad \bot \quad \gamma^{\eqref{kktDAM_w_lim1}}_j \geq 0, && \quad \forall j \in \mathcal{J}, \\
& 0 \geq  w_j - \widehat{W}_j \quad \bot \quad \gamma^{\eqref{kktDAM_w_lim2}}_j \geq 0, && \quad \forall j \in \mathcal{J}, \\ 
\begin{split}
    & 0 \geq \boldsymbol{M}_{(\ell,\cdot)} \left( \boldsymbol{H}_\text{G}^\top \boldsymbol{p} + \boldsymbol{H}_\text{J}^\top \boldsymbol{w} - \boldsymbol{D} \right)  - F_\ell \\
    & \quad \quad \quad \quad  \bot \quad \gamma^{\eqref{kktDAM_f_lim1}}_\ell \geq 0, 
\end{split} && \quad  \forall \ell \in \mathcal{L}, \\
\begin{split}
    & 0 \geq - F_\ell - \boldsymbol{M}_{(\ell,\cdot)} \left( \boldsymbol{H}_\text{G}^\top \boldsymbol{p} + \boldsymbol{H}_\text{J}^\top \boldsymbol{w} - \boldsymbol{D} \right)  \\
    & \quad \quad \quad \quad  \bot \quad \gamma^{\eqref{kktDAM_f_lim2}}_\ell \geq 0, 
\end{split} && \quad  \forall \ell \in \mathcal{L}. \label{KKT_DA_last}
\end{align}


\section{}\label{app:B}
A multi-cut Bender's decomposition scheme is implemented \cite{Conejo2006}. The master problem at iteration $\eta$ is formulated as
\begin{mini!}|s|[2]                   
    {_{\Xi_\text{MP}}} 
    { \mathcal{C}_\text{R} + \mathcal{C}_\text{D} + \sum_{s \in \mathcal{S}} \pi_s   \vartheta_s }{\label{prob:MP_Bender}}
    {} 
    \addConstraint{ \text{\eqref{KKT_RMz_1} - \eqref{KKT_RMz_last}},\nonumber}{}{ \hspace{-5cm} \text{KKT of zonal reserve market}  }
    \addConstraint{ \text{\eqref{KKT_DA_1} - \eqref{KKT_DA_last}},\nonumber}{}{ \hspace{-5cm} \text{KKT of day-ahead market}  }
    \addConstraint{ \eqref{con:ResReq}, \nonumber}{}{ \hspace{-5cm}  \text{Zonal reserve requirements}, }
    \addConstraint{ \eqref{eq:part_xn} - \eqref{eq:part_cz}, \nonumber}{}{ \hspace{-5cm} \text{Grid partitioning}, }
    \addConstraint{ \eqref{eq:h} - \eqref{eq:gamma2}, \nonumber}{}{ \hspace{-5cm}  \text{Capacity allocation}, }
    \addConstraint{ \vartheta_s \geq \vartheta_0  }{}{\hspace{-5cm} \forall s \in \mathcal{S}}
    \addConstraint{ \vartheta_s \geq \mathcal{C}_{\text{B},s}^{(k)} + \sum_{g} \gamma^{\eqref{con:alpha}^{(k)}}_{g,s} \left( r_g^+ - r_g^{+;(k)} \right) \ldots }
        \addConstraint{ \hspace{-1cm} + \sum_{g} \gamma^{\eqref{con:beta}^{(k)}}_{g,s} \left( r_g^- - r_g^{-;(k)} \right) + \sum_{g} \gamma^{\eqref{con:gamma}^{(k)}}_{g,s} \left( p_g - p_g^{(k)} \right) \nonumber \ldots }
        \addConstraint{ \hspace{-1cm} + \sum_{j} \gamma^{\eqref{con:delta}^{(k)}}_{j,s} \left( w_j - w_j^{(k)} \right), \, \forall s \in \mathcal{S}, \, \forall k = 1,\ldots,\eta-1  \nonumber}
\end{mini!}
where $\gamma^{(*)^{(k)}}$ are the dual variables of constraints $(*)$ in the sub-problem, whose formulation for scenario $s=s'$ and iteration $\eta$ is the following
\begin{mini!}|s|[2]                   
    {_{\Xi_\text{SP}}} 
    { \mathcal{C}_{\text{B},s'} = \sum_{g \in \mathcal{G}} C_g \left( p_{g,s'}^+ - p_{g,s'}^- \right)\ldots \\ + \sum_{j \in \mathcal{J}} C^\text{ct} w_{j,s'}^\text{ct}  + \sum_{n \in \mathcal{N}} C^\text{sh} d_{n,s'}^\text{sh} \nonumber  }{\label{prob:SP_Benders}} 
    {} 
    {}{}    
    \addConstraint{ \hspace{-4.5cm} r_g^+ = r_g^{+;(\eta)}}{}{ \hspace{+0.5cm} \hspace{-3cm} :\gamma^{\eqref{con:alpha}^{(\eta)}}_{g,s'}  \label{con:alpha} }
    \addConstraint{ \hspace{-4.5cm} r_g^- = r_g^{-;(\eta)}}{}{ \hspace{+0.5cm} \hspace{-3cm} :\gamma^{\eqref{con:beta}^{(\eta)}}_{g,s'} \label{con:beta}}
    \addConstraint{ \hspace{-4.5cm} p_g = p_g^{(\eta)}}{}{ \hspace{+0.5cm} \hspace{-3cm} :\gamma^{\eqref{con:gamma}^{(\eta)}}_{g,s'} \label{con:gamma} }
    \addConstraint{ \hspace{-4.5cm} w_j^+ = w_j^{(\eta)}}{}{ \hspace{+0.5cm} \hspace{-3cm} :\gamma^{\eqref{con:delta}^{(\eta)}}_{j,s'} \label{con:delta} }
    \addConstraint{ \hspace{-4.5cm}\eqref{BM_bal} - \eqref{BM_f} \nonumber, }{}{ \hspace{+0.5cm} \hspace{-3cm} \text{Balancing market}, \quad s = s',}
\end{mini!}
where first-stage decision variables $r_g^+$, $r_g^-$, $p_g$ and $w_j$ are fixed to the solution of the master problem at the current iteration. As the problem has complete recourse, no need for feasibility cuts is required and $\mathcal{S}$ new optimality cuts are added to the master problem at each iteration. The algorithm converges to a solution when the condition $ | \sum_{s} \pi_s \vartheta_s - \sum_{s} \pi_s \mathcal{C}_{\text{B},s}^{(\eta)} | \leq \varepsilon $ is satisfied for a predefined tolerance $\varepsilon$.


%% file: 00_Main.bbl
\begin{thebibliography}{10}
\providecommand{\url}[1]{#1}
\csname url@samestyle\endcsname
\providecommand{\newblock}{\relax}
\providecommand{\bibinfo}[2]{#2}
\providecommand{\BIBentrySTDinterwordspacing}{\spaceskip=0pt\relax}
\providecommand{\BIBentryALTinterwordstretchfactor}{4}
\providecommand{\BIBentryALTinterwordspacing}{\spaceskip=\fontdimen2\font plus
\BIBentryALTinterwordstretchfactor\fontdimen3\font minus
  \fontdimen4\font\relax}
\providecommand{\BIBforeignlanguage}[2]{{%
\expandafter\ifx\csname l@#1\endcsname\relax
\typeout{** WARNING: IEEEtran.bst: No hyphenation pattern has been}%
\typeout{** loaded for the language `#1'. Using the pattern for}%
\typeout{** the default language instead.}%
\else
\language=\csname l@#1\endcsname
\fi
#2}}
\providecommand{\BIBdecl}{\relax}
\BIBdecl

\bibitem{Morales2009}
{J.M. Morales, A.J. Conejo and J. Pérez-Ruiz}, ``{Economic Valuation of
  Reserves in Power Systems With High Penetration of Wind Power},'' \emph{IEEE
  Trans. on Power Syst.}, vol.~24, no.~2, pp. 900--910, 2009.

\bibitem{Papavasiliou2013}
A.~Papavasiliou and S.~S. Oren, ``Multiarea stochastic unit commitment for high
  wind penetration in a transmission constrained network.'' \emph{Operations
  Research}, vol.~61, no.~3, pp. 578 -- 592, 2013.

\bibitem{Morales2014}
J.~M. Morales, M.~Zugno, S.~Pineda, and P.~Pinson, ``{Electricity market
  clearing with improved scheduling of stochastic production},'' \emph{European
  Journal of Operational Research}, vol. 235, no.~3, pp. 765--774, 2014.

\bibitem{Dvorkin2019}
{V. Dvorkin, S. Delikaraoglou, J.M. Morales}, ``{Setting reserve requirements
  to approximate the efficiency of the stochastic dispatch},'' \emph{IEEE
  Trans. on Power Syst.}, vol.~34, no.~2, pp. 1524--1536, 2019.

\bibitem{Delikaraoglou2019}
{S. Delikaraoglou, P. Pinson}, ``{Optimal allocation of HVDC interconnections
  for exchange of energy and reserve capacity services},'' \emph{Energy
  Systems}, vol.~10, no.~3, pp. 635--675, 2019.

\bibitem{Fan2012}
N.~Fan, D.~Izraelevitz, F.~Pan, P.~M. Pardalos, and J.~Wang, ``{A mixed integer
  programming approach for optimal power grid intentional islanding},''
  \emph{Energy Systems}, vol.~3, no.~1, pp. 77--93, 2012.

\bibitem{Golari2014}
M.~Golari, N.~Fan, and J.~Wang, ``{Two-stage stochastic optimal islanding
  operations under severe multiple contingencies in power grids},''
  \emph{Electric Power Systems Research}, vol. 114, pp. 68--77, 2014.

\bibitem{Kumar2004}
A.~{Kumar}, S.~C. {Srivastava}, and S.~N. {Singh}, ``A zonal congestion
  management approach using real and reactive power rescheduling,'' \emph{IEEE
  Trans. on Power Syst.}, vol.~19, no.~1, pp. 554--562, Feb 2004.

\bibitem{Wang2012}
F.~{Wang} and K.~W. {Hedman}, ``Reserve zone determination based on statistical
  clustering methods,'' in \emph{2012 North American Power Symposium (NAPS)},
  Sep. 2012, pp. 1--6.

\bibitem{Wang2015}
F.~Wang and K.~W. Hedman, ``{Dynamic reserve zones for day-ahead unit
  commitment with renewable resources},'' \emph{IEEE Trans. on Power Syst.},
  vol.~30, no.~2, pp. 612--620, 2015.

\bibitem{Chen2014}
Y.~Chen, P.~Gribik, and J.~Gardner, ``{Incorporating Post Zonal Reserve
  Deployment Transmission Constraints Into Energy and Ancillary Service
  Co-Optimization},'' \emph{IEEE Trans. on Power Syst.}, vol.~29, no.~2, pp.
  537--549, 2014.

\bibitem{Khatir2013}
A.~{Ahmadi-Khatir}, M.~{Bozorg}, and R.~{Cherkaoui}, ``Probabilistic spinning
  reserve provision model in multi-control zone power system,'' \emph{IEEE
  Trans. on Power Syst.}, vol.~28, no.~3, pp. 2819--2829, Aug 2013.

\bibitem{Gebrekiros2013}
Y.~T. Gebrekiros, G.~L. Doorman, H.~Farahmand, and S.~Jaehnert, ``{Benefits of
  cross-border reserve procurement based on pre-allocation of transmission
  capacity},'' \emph{IEEE Grenoble Conf. PowerTech}, pp. 1--6, 2013.

\bibitem{Trespalacios2014}
F.~Trespalacios and I.~Grossmann, ``Review of mixed-integer nonlinear and
  generalized disjunctive programming methods,''
  \emph{Chemie-Ingenieur-Technik}, vol.~86, no.~7, pp. 991--1012, 2014, cited
  By 72.

\bibitem{Pineda2019}
S.~{Pineda} and J.~M. {Morales}, ``Solving linear bilevel problems using
  big-ms: Not all that glitters is gold,'' \emph{IEEE Transactions on Power
  Systems}, vol.~34, no.~3, pp. 2469--2471, 2019.

\bibitem{Conejo2006}
A.~J. Conejo, E.~Castillo, R.~M{\'{i}}nguez, and R.~Garc{\'{i}}a-Bertrand,
  \emph{{Decomposition techniques in mathematical programming: Engineering and
  science applications}}.\hskip 1em plus 0.5em minus 0.4em\relax Springer,
  2006.

\bibitem{Bludszuweit2008}
H.~Bludszuweit, J.~A. Dom{\'{i}}nguez-Navarro, and A.~Llombart, ``{Statistical
  analysis of wind power forecast error},'' \emph{IEEE Trans. on Power Syst.},
  vol.~23, no.~3, pp. 983--991, 2008.

\bibitem{Papaefthymiou2009}
G.~Papaefthymiou and D.~Kurowicka, ``{Using copulas for modeling stochastic
  dependence in power system uncertainty analysis},'' \emph{IEEE Trans. on
  Power Syst.}, vol.~24, no.~1, pp. 40--49, 2009.

\bibitem{Dupacova2000}
J.~Dupa{\v{c}}ov{\'{a}}, G.~Consigli, and S.~W. Wallace, ``{Scenarios for
  Multistage Stochastic Programs},'' \emph{Annals of Operations Research}, vol.
  100, no. 1-4, pp. 25--53, 2000.

\bibitem{rts24}
C.~Ordoudis, P.~Pinson, and M.~Zugno, ``An updated version of the {IEEE} rts
  24-bus system for electricity market and power system operation studies,''
  \emph{Technical University of Denmark (DTU)}, pp. pp. 1--5, 2016.

\end{thebibliography}
